\documentclass[prb,superscriptaddress,longbibliography, twocolumn]{revtex4-1}

\usepackage{times,amsmath}
\usepackage{epsfig}
\usepackage{color}
\usepackage{longtable}
\usepackage{ulem}
\usepackage{graphicx}
\usepackage{dcolumn}
\usepackage{bm}
\usepackage{bookmark}
\usepackage{tabularx}
\usepackage{hyperref}
\usepackage{multirow}

\hypersetup{colorlinks=true, citecolor=blue, filecolor=blue, linkcolor=blue, urlcolor=blue}

\begin{document}

\title{Bending Deformation Driven by Molecular Rotation}

\author{Pedro A. Santos-Florez}
\affiliation{Department of Physics and Astronomy, University of Nevada, Las Vegas, NV 89154, USA}

\author{Shinnosuke Hattori}
\affiliation{Advanced Research Laboratory, Technology Infrastructure Center, Technology Platform, Sony Group Corporation, 4-14-1 Asahi-cho, Atsugi-shi 243-0014, Japan}

\author{Qiang Zhu}
\email{qzhu8@uncc.edu}
\affiliation{Department of Physics and Astronomy, University of Nevada, Las Vegas, NV 89154, USA}
\affiliation{Department of Mechanical Engineering and Engineering Science, University of North Carolina at Charlotte, Charlotte, NC 28223, USA}

\date{\today}

\begin{abstract}
In recent years, certain molecular crystals have been reported to possess surprising flexibility by undergoing significant elastic or plastic deformation in response to mechanical loads. However, despite this experimental evidence, there currently exists no atomistic mechanism to explain the physical origin of this phenomenon from numerical simulations. In this study, we investigate the mechanical behavior of three naphthalene diimide derivatives, which serve as representative examples, using direct molecular dynamics simulations. Our simulation trajectory analysis suggests that molecular rotational freedom is the key factor in determining a crystal's mechanical response, ranging from brittle fracture to elastic or plastic deformation under mechanical bending. Additionally, we propose a rotation-dependent potential energy surface as a means to classify organic materials' mechanical responses and identify new candidates for future investigation.
\end{abstract}


\vskip 300 pt

\maketitle

\section{INTRODUCTION}
\begin{figure}[htbp]
\centering
\includegraphics[width=0.5 \textwidth]{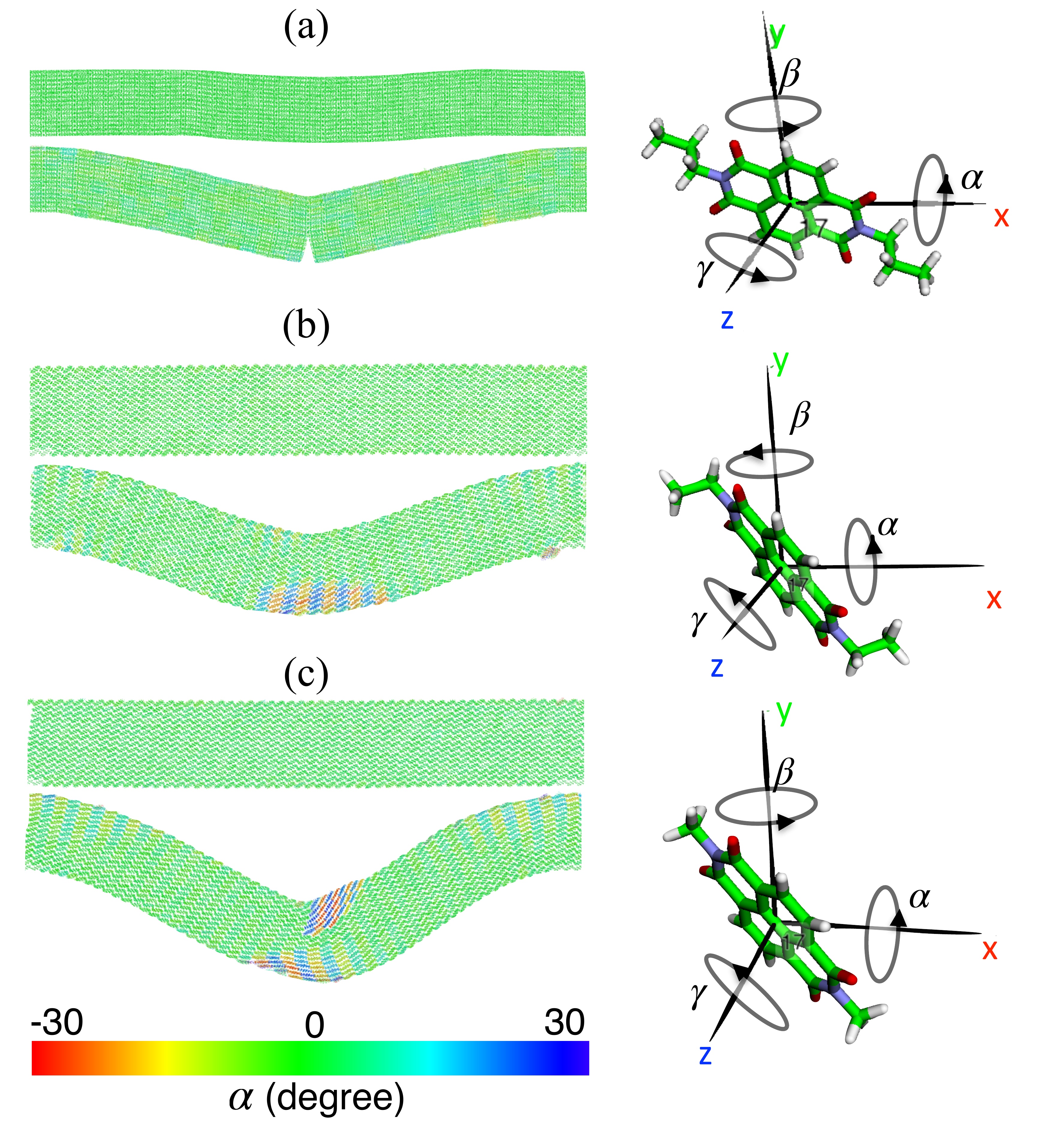}
\caption{\label{Fig1} The simulated bending on three different materials based on naphthalene diimide derivatives. (a) brittle \textbf{Pr} (50.3$\times$7.0$\times$6.8 nm$^3$), (b) elastic \textbf{Et} (50.7$\times$6.4$\times$6.6 nm$^3$) and (c) elastic/plastic \textbf{Me} (50.2$\times$6.4$\times$6.9 nm$^3$). These three crystals consist of very similar molecules that differ only in the side groups. In the left panel, the initial and finally deformed configurations are colored by the accumulated molecular rotation ($\alpha$) along the $x$-axis. The corresponding molecules and the definition of rotation angles are shown in the right panel.} 
\vspace{-5mm}
\end{figure}

While most molecular crystals are brittle, there exists a class of compliant organic crystals that can bend under a large mechanical stress loading \cite{naumov2015mechanically, saha2018molecules}. Since early 2000, a growing number of mechanically flexible organic crystals have been reported experimentally \cite{reddy2006structure, Takamizawa2013Superelastic, ghosh2012elastic, panda2015spatially, krishna2016mechanically, yadav2019molecular, raju2018rationalizing, mishra2020conformation, zhang2021structural}. In general, the mechanical response of an organic solid depends on both the molecular substance and crystal packing. A remarkable example is shown in Fig. \ref{Fig1}. Three crystals, made of similar molecules from naphthalene diimide derivatives, were found to exhibit distinct responses from brittle fracture to compliant deformation with either reversible (elastic) or irreversible (plastic) characteristic \cite{devarapalli2019remarkably}. The flexible nature of organic materials is vital for a variety of applications, e.g., high-performance modular solar cells \cite{root2017mechanical}, actuators \cite{li2019martensitic}, photochemistry \cite{mutai2021organic}, fluorescence \cite{di2022fluorescence, hayashi2016elastic}, electronics \cite{wang2019organic, samanta2023elastic}, optics \cite{liu2018highly}, as well as drug tabulation \cite{sun2017microstructure}. 

In the recent years, various computational techniques have been introduced to characterize the mechanical properties of molecular systems \cite{devarapalli2019remarkably, wang2018identifying, wang2019computational, wang2020landscape, ootani2022density, matveychuk2022quantitative}. They include topological analysis, elastic properties calculation \cite{wang2019computational}, and shear/tensile simulations \cite{devarapalli2019remarkably, ootani2022density}. These techniques are successful in identifying brittle materials. Within an interlocked environment in crystal packing, molecular motions in those materials are largely restricted, resulting a brittleness under bending \cite{wang2019computational}. On the other hand, some materials are featured by a strong anisotropy with plausible slip planes \cite{wang2019computational, bryant2018predicting}. Therefore, these materials become compliant over a broad range of applied stress along some crystallographic directions. However, all available techniques fail to explain the difference between the elastic and plastic materials. While there have been plenty of studies on the bending of metals \cite{ZHU20127112, Zhang_2014, NOHRING2016140, ZHUO2018331, KATAKAM2020106674, HE20223687}, to our knowledge, no attempts have been made to directly simulate the bending of organic materials at the atomistic level.

Among the compliant crystals, ductile materials are often favored in engineering applications \cite{sun2017microstructure}. Hence, researchers attempted to use the well established dislocation theory to explain the observed plasticity on organic materials \cite{reddy2006structure, saha2018molecules}. Similar to the plastic deformation in ductile metals, it was proposed that mechanical shearing can occur via the slippage of dislocated molecular layers on the molecular crystals with a layered packing \cite{reddy2010mechanical, wang2018identifying}. Using these slip planes, a bending model was proposed to explain the underlying mechanism \cite{reddy2005structural}. Although the dislocation is not uncommon in molecular crystals \cite{mathew2013peierls, mathew2013slip, olson2018dislocations}, there has been no direct experimental evidence to support that dislocation pre-exists or appears in the organic crystals under bending. Furthermore, this mechanism fails to explain the observed large-scale elastic deformation. In fact, two crystals in Fig. \ref{Fig1}b-c can undergo either elastic or plastic deformation despite the apparent similarity. Clearly, our current understanding on molecular bending remains limited. 

In this work, we present our efforts in questing the molecular bending mechanism through atomistic simulation. To achieve this goal, we started by developing a simulation protocol that can directly model the bending of organic crystals at the atomic level. Next, the simulation results were carefully analyzed to classify and understand the atomistic mechanism of materials-dependent deformations from brittle fracture to elastic or plastic deformation. Furthermore, we demonstrated that molecular rotational freedom is the key factor in determining a crystal's mechanical response. Finally, we introduced a rotation-dependent potential energy surface as a means to classify organic materials' mechanical responses and identify new candidates for future screening of new mechanically flexible organic crystals.

\section{Methods}

\subsection{Crystal Structures of Three Systems}
\begin{figure}[htbp]
\centering
\includegraphics[width=0.50\textwidth]{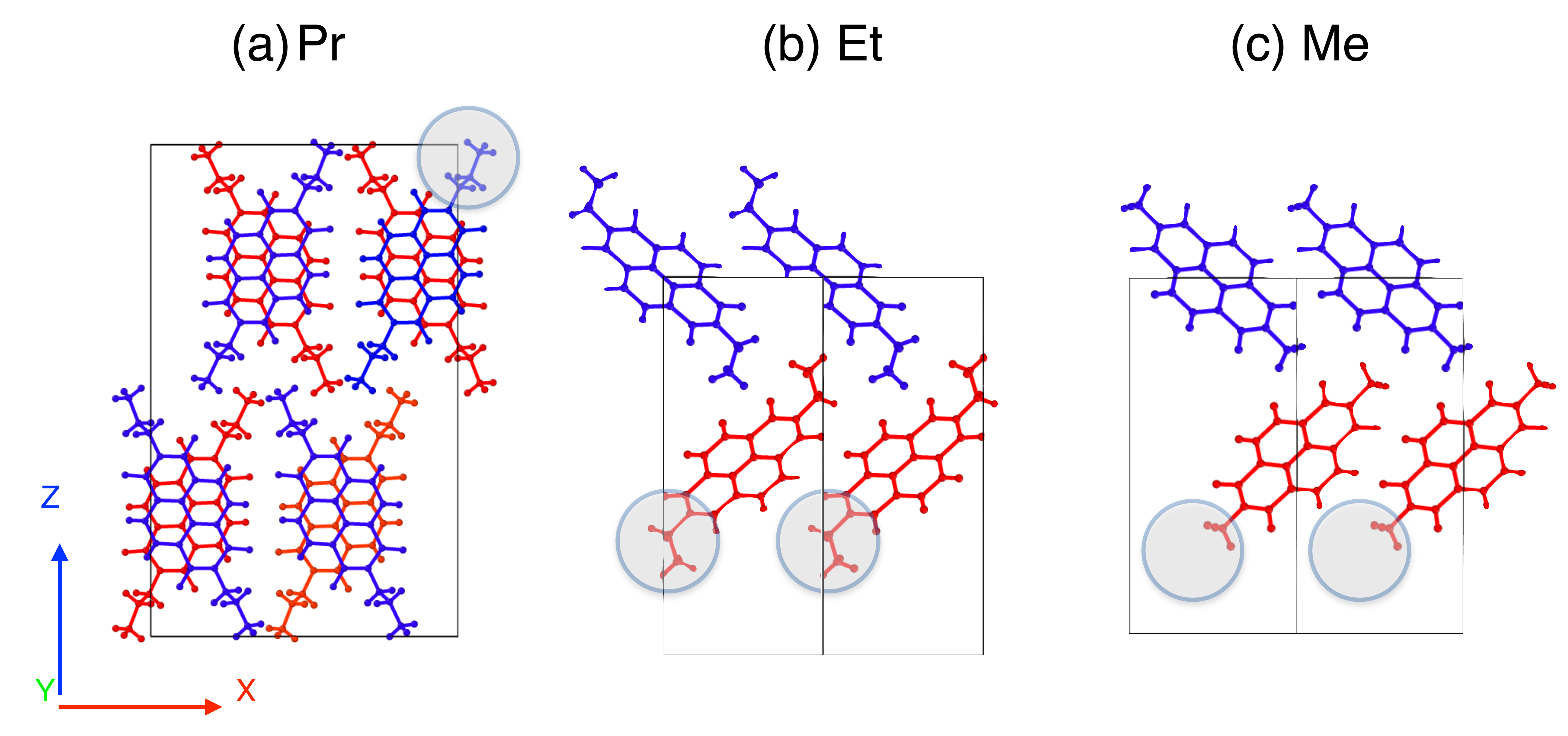}
\caption{\label{Fig2} The crystal structures of (a) Pr, (b) Et (c) Me systems.}
\end{figure}

In this study, we focused on three systems consisting of naphthalene diimide derivatives as shown in Fig. \ref{Fig1}. The three molecules share the same backbone while differing only in the side chains. The brittle crystal consists of the molecules with the \textbf{propyl} group, featured by the orthorhombic space group $Pbca$ with one molecule in the asymmetric unit. On the other hand, the elastic/plastic crystals have the \textbf{ethyl/methyl} groups, both adopting the monoclinic space group $P2_1/c$ with half a molecule in the asymmetric unit. For convenience, we follow the previous literature \cite{devarapalli2019remarkably} to name these systems according to their molecular functional groups (i.e., \textbf{Pr, Et, Me}). In all three cases, the weak interaction plane formed by alkyl groups is (001). In Fig. \ref{Fig2}, each molecule in the unit cell is colored by the alignment along the $y$-axis. Clearly, the overall molecular packing in the brittle-\textbf{Pr} crystal are more complex. Since there exist eight different types of molecular alignments due to the $mmm$ symmetry operations, the \textbf{Pr} crystal has molecules aligned in different ways within the same (001) layer. On the contrary, there are only two types of molecular alignments in the \textbf{Et}/\textbf{Me} crystals. And the (001) layer in \textbf{Et}/\textbf{Me} crystals has all molecules aligned in the same direction. Table \ref{table1} summarizes the crystallographic information of three molecular crystals.  

\begin{table}[ht]
  \centering
  \caption{The crystallographic information of three molecular crystals. Among them, \textbf{Pr} denotes the brittle crystal with the CSD refcode of DAHLOQ; \textbf{Et} is the elastic crystal with the CSD refcode of BIYRIM01; and \textbf{Me} is the plastic crystal with the CSD refcode of DAHMUX. The column of size list the number of molecules in each model.}
\begin{tabular}{cccccccc}
\hline\hline
System&~~Symmetry~~&  ~~Size~~ &~~$a$~(\AA)~~&~~$b$~(\AA)~~&~~$c$~(\AA)~~&~~$\beta$~($^\circ$)~~\\ \hline
Pr     & $Pbca$        & 8                   & 6.96  & 17.24  & 27.58 & 90.0  \\ 
Et     & $P2_1/c$       & 2                  & 4.84  & 7.74   & 18.32 & 90.1  \\ 
Me     & $P2_1/c$       & 2                  & 4.62  & 8.02   & 17.02 & 94.0  \\ 
\hline\hline
\label{table1}
\end{tabular}
\end{table}

\subsection{Atomistic Modeling of Bending}
\begin{figure*}[htbp]
\centering
\includegraphics[width=0.90 \textwidth]{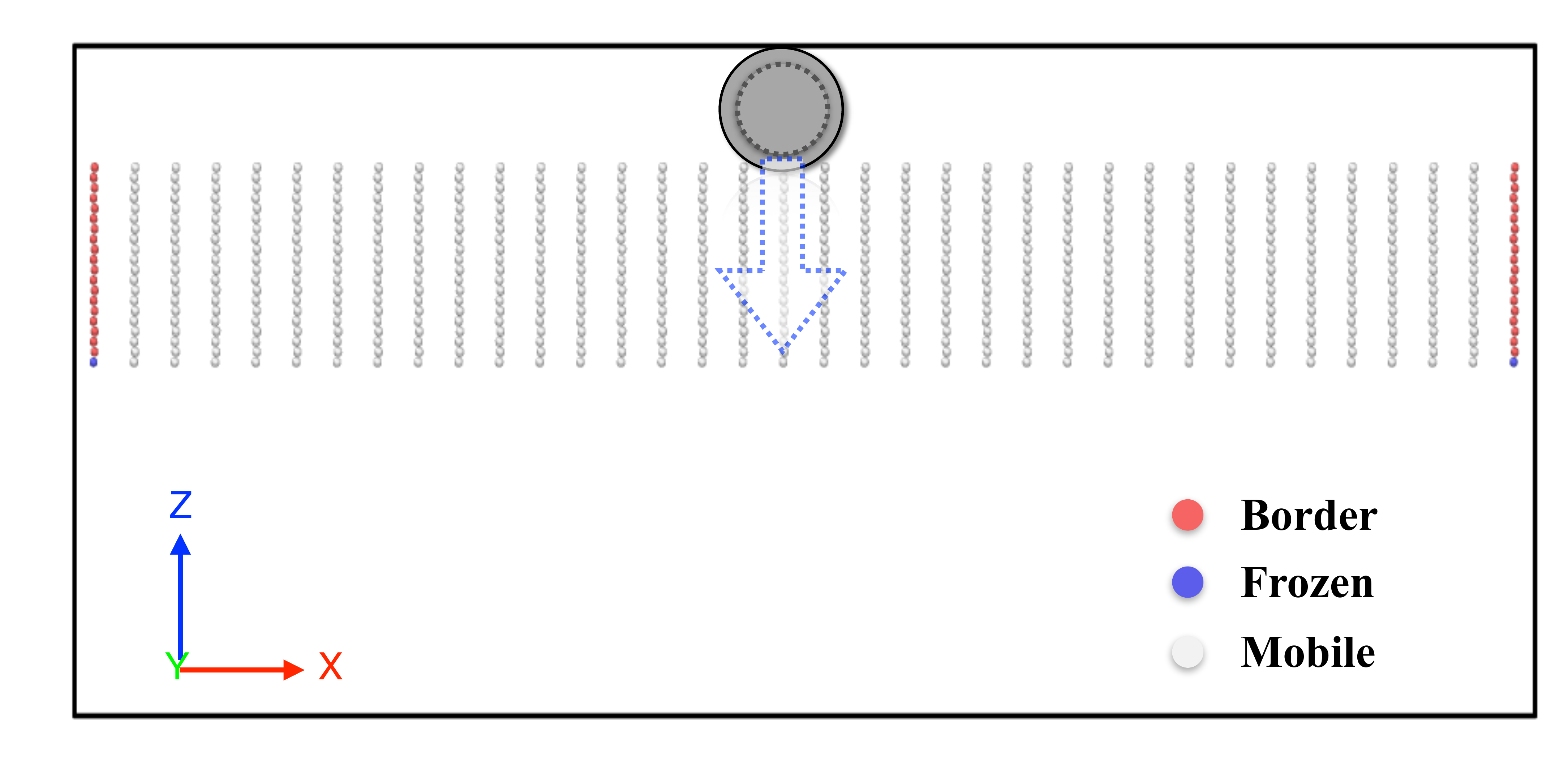}
\caption{\label{Fig3} The schematic setup of a bending simulation model.}
\end{figure*}

To directly simulate the bending of organic crystal, we employed a three-point bending model within a partial periodic boundary condition based on the \texttt{LAMMPS} package \cite{lammps} at room temperature. In our calculation, we performed non-equilibrium molecular dynamics (MD) simulation by applying the indentation on the molecular slab model (see Fig. \ref{Fig3}). Both $x$ and $y$-axes are under the constraint of periodic boundary conditions, while the $c$-axis is not periodic. We rotated the crystal structures with the matrix of [[0,0,1], [0,-1,0], [1,0,0]], and then built the super cell slab models with sufficient vacuum separation. The slab correction was applied to remove the slab-slab interactions from the periodic images. Due to the non-triclinic box restriction on the computation of slab correction, the $\beta$ angles for the slabs of \textbf{Et} and \textbf{Me} were to be set to 90$^\circ$, which are slightly different from the ideal values. However, this compromise should not change the results largely. 

Along the non-periodic $z$-axis, a cylindrical indenter with the radius of 30 \AA~ is applied on top of the slab center in the unit cell. To mimic two other contacting points in the three-points bending simulation, the last layer of molecules in the bottom region were \textit{frozen} in the entire simulation. In addition, the first columns of molecules on both left and right sides of the unit cell are defined as the \textit{border}. The rest atoms not belonging to the \textit{frozen} and \textit{border} groups are set to the \textit{mobile} group that can move freely. To ensure a sufficient heat bath, we first performed Langevin thermostat \cite{langevin} on both \textit{mobile} and \textit{border} groups, followed by a second thermal equilibration on only the \textit{border} atoms. The fully equilibrated sample was used to perform three-points bending simulation with only the \textit{border} atoms being under the Langevin thermostat to mimic the external temperature reservoir. Upon bending, an indenter was used to push into the simulation slab in a flow with the rate of 10 m/s. When the system reaches the maximum indentation depth, the indenter was kept for 300-500 ps to allow the system achieves thermal equilibrium. Afterwards, the indenter will move upward with the previous rate to mimic the release of indenter process. To check the dependence of indentation rate, we also varied the rates from 2-50 m/s. We found that these rates roughly led to similar results. However, a rate faster than 200 m/s may result in nonphysical phase transition for the \textbf{Me} sample. It is also possible that the change of indenter shape, size and temperature may change the results significantly. These factors will be the subject of future work.

\subsection{Force Field Choices and Benchmark}
To reliably simulate the deformation of organic crystals at the atomistic level, it is crucial to choose an accurate interatomic force field model. In this work, we developed a computational pipeline to automate the generation of molecular force fields from \texttt{AmberTools20} \cite{Case2020a}, based on the General Amber Force Field (GAFF) \cite{gaff} framework with atomic charges using semi-empirical AM1-BCC method \cite{jakalian2000fast}. To confirm that the simulation results are not due to the artifact of force field choices, we repeated the simulations using the \texttt{OpenFF-toolkit} \cite{jeff_wagner_2023_7506404} with different parametrization protocol \cite{openff2.0.0, parmed}, as well as Density Functional based Tight Binding (DFTB) \cite{dftb+}. The \texttt{OpenFF-toolkit}\cite{jeff_wagner_2023_7506404} was employed to generate the OpenFF model by assigning atom types based on direct chemical perception, utilizing an atom-by-atom assignment approach through the use of SMIRKS (SMILES Arbitrary Target Specification) patterns. The OpenFF 2.0.0 (Sage) force field \cite{openff2.0.0} was adopted for atom typing, and the \texttt{ParmEd} package \cite{parmed} was employed for input file format conversion, ensuring compatibility with various molecular simulation engines and the accurate representation of molecular topologies. In the DFTB model, we used the \texttt{DFTB+} code \cite{dftb+} with the inclusion of van de Waals dispersion based on the Tkatchenko-Scheffler method \cite{TS-PRL}. 

Table \ref{table2} lists the computed equilibrium cell parameters with different types of force fields at both zero and room temperatures. Clearly, the GAFF model, as well as other models, yield similar results that are comparable with the experimental values \cite{devarapalli2019remarkably}. For convenience, we will mainly employ the GAFF model in our following simulations.

\begin{table}[ht]
  \centering
  \caption{The comparison of different force field models in describing the equilibrium cell parameters. }
\begin{tabular}{ccccccccc}
\hline\hline
System& ~Cell~ &~Experiment~ & ~CVFF~ & ~GAFF~ & ~OpenFF~ & ~DFTB~ \\ 
              & (\AA)      & 300 K         &  300 K     & 300 K       & 300 K     & 0 K \\ 
              &    & Ref. \cite{devarapalli2019remarkably} & Ref. \cite{devarapalli2019remarkably}& this work& this work &  this work \\ \hline
\multirow{4}{*}{Pr}  & $a$      & 6.96  & 7.54     & 7.30  &  7.35  &  6.69  \\
                     & $b$      & 17.24 & 16.99    & 17.41 &  17.62 &  17.08 \\ 
                     & $c$      & 27.58 & 28.14    & 27.09 &  27.20 &  27.74 \\ \hline
\multirow{4}{*}{Et}  & $a$      & 4.84  & 5.02     & 5.07  &  4.93  &  4.53  \\  
                     & $b$      & 7.74  & 7.66     & 7.79  &  7.86  &  7.81  \\ 
                     & $c$      & 18.32 & 19.88    & 19.07 &  19.05 & 18.52  \\ \hline
\multirow{4}{*}{Me}  & $a$      & 4.62  & 4.60     & 4.58  & 4.50   &  4.29  \\ 
                     & $b$      & 8.02  & 7.87     & 8.28  & 8.19   &  8.02  \\ 
                     & $c$      & 17.02 & 18.66    & 18.40 & 17.82  &  16.69 \\ \hline\hline
\label{table2}
\end{tabular}
\end{table}

\section{Computational Results and Discussions}
To make a fair comparison, we set up all model sizes close to 50.0 $\times$ 7.0 $\times$ 7.0 nm$^3$ as summarized in Table \ref{table3}. For each system, we also added the vacuum of 120~\AA~to allow the materials bend sufficiently. In addition, we considered two kinds of \textbf{Me} models, including (i) the supercell after the isobaric-isothermal (NPT) equilibration; and (ii) the supercell with the experimental cell parameters. Although these two configurations only differ slightly, it has been found they led to different elastic/plastic deformation processes in the subsequent bending simulation. All supercell slab models were then used to perform the three-points bending simulation as illustrated in Fig. \ref{Fig3} with an indentation rate of 10 m/s under 300 K. For each system, we ran the indentation simulation for multiple times to determine the maximum indentation depth (5-20 nm) that leads to the formation of crack. Before the maximum indentation depth is reached, we also continued the simulation by releasing the indenter with the same rate to check if the deformation process is reversible.

\begin{table}[ht]
  \centering
  \caption{The details of slab models used in the bending simulation. The column of size lists the number of molecules in each model.}
\begin{tabular}{ccccccc}
\hline\hline
System &~Deformation~& ~~~Supercell~~~ & Size & ~~~$a$~(\AA)~~&~~$b$~(\AA)~~&~~$c$~(\AA)~~\\ \hline
Pr         & brittle    & $18\times4\times5$ & 5760 & 503.2 & 69.9 & 70.6  \\ 
Et         & elastic    & $27\times4\times5$ & 6480 & 508.5 & 63.6 & 74.7  \\ 
Me         & elastic    & $29\times8\times15$ & 6960 & 501.6 & 65.2 & 86.5  \\ 
Me         & plastic    & $30\times8\times15$ & 7200 & 510.6 & 64.2 & 85.1  \\ 
\hline\hline
\label{table3}
\end{tabular}
\end{table}

\subsection{Direct Identification of Deformation Characteristics}
\begin{figure}[ht]
\centering
\includegraphics[width=0.5 \textwidth]{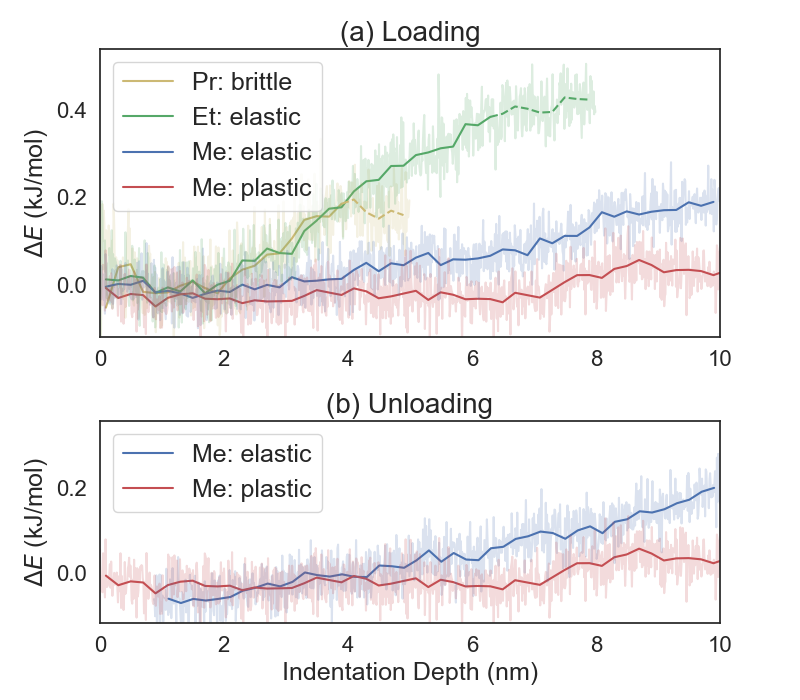}
\caption{\label{Fig4} The evolution of average molecular potential energy as a function of indentation depth upon (a) loading and (b) unloading. 
}
\end{figure}

Fig. \ref{Fig4} summarizes the simulated evolution of potential energy as a function of indentation depth for all three materials. Encouragingly, our calculations produced a sequence of deformations (including brittle fracture, elastic deformation and plastic bending) that are similar to the previous experimental observations \cite{devarapalli2019remarkably}. First, \textbf{Pr} is clearly brittle as evidenced by the abrupt drop of energy in Fig. \ref{Fig4}a, which is consistent to the appearance of crack pattern in Fig. \ref{Fig1}a when the indenter reaches 3.5 nm. On the other hand, \textbf{Et} is more complaint with a maximum indentation of 6.2 nm. Applying further loading would lead to the formation of crack as well. If we release the indentation before \textbf{Et} reaches 6.2 nm, the model roughly returned to the original state. Therefore, this deformation is elastic. Interestingly, \textbf{Me} can survive under more than 10 nm indentation without breaking under two different setups. For the slab after a full NPT equilibration, it bends elastically, as evidenced by the reversible energy versus indentation depth relation (denoted as \textbf{Me}-elastic in Fig. \ref{Fig4}b). When the slab has a small strain in the initial configuration (see Table \ref{table3}), its energy curves upon loading and unloading are no longer reversible. Compared to \textbf{Me}-elastic, this sample achieves a lower energy when it approaches the maximum indentation depth upon loading. When the indentation is released, it does not return to the original state, but maintains a relatively higher energy. Therefore, the whole deformation process is irreversible and plastic. The sample will be called \textbf{Me}-plastic from now on. 

To our knowledge, all previous computational studies were limited to indirect simulations of tensile and shear tests \cite{devarapalli2019remarkably, wang2019computational, ootani2022density,  matveychuk2022quantitative}. Here, our calculations provide the first direct atomistic modeling on the experimentally observed bending deformations. Compared to the simulation results, the elastic and plastic samples are found to bend more significantly in real experiments \cite{devarapalli2019remarkably}. This is because that the material along $x$-axis under the actual bending test can shrink to release the tensile stress. However, our simulation model still obeys the periodic boundary condition along the $x$-axis. Hence we expect that the degree of bending in simulation is underestimated as compared to the real situation. 

\begin{figure}[ht]
\centering
\label{Fig5}
\includegraphics[width=0.5 \textwidth]{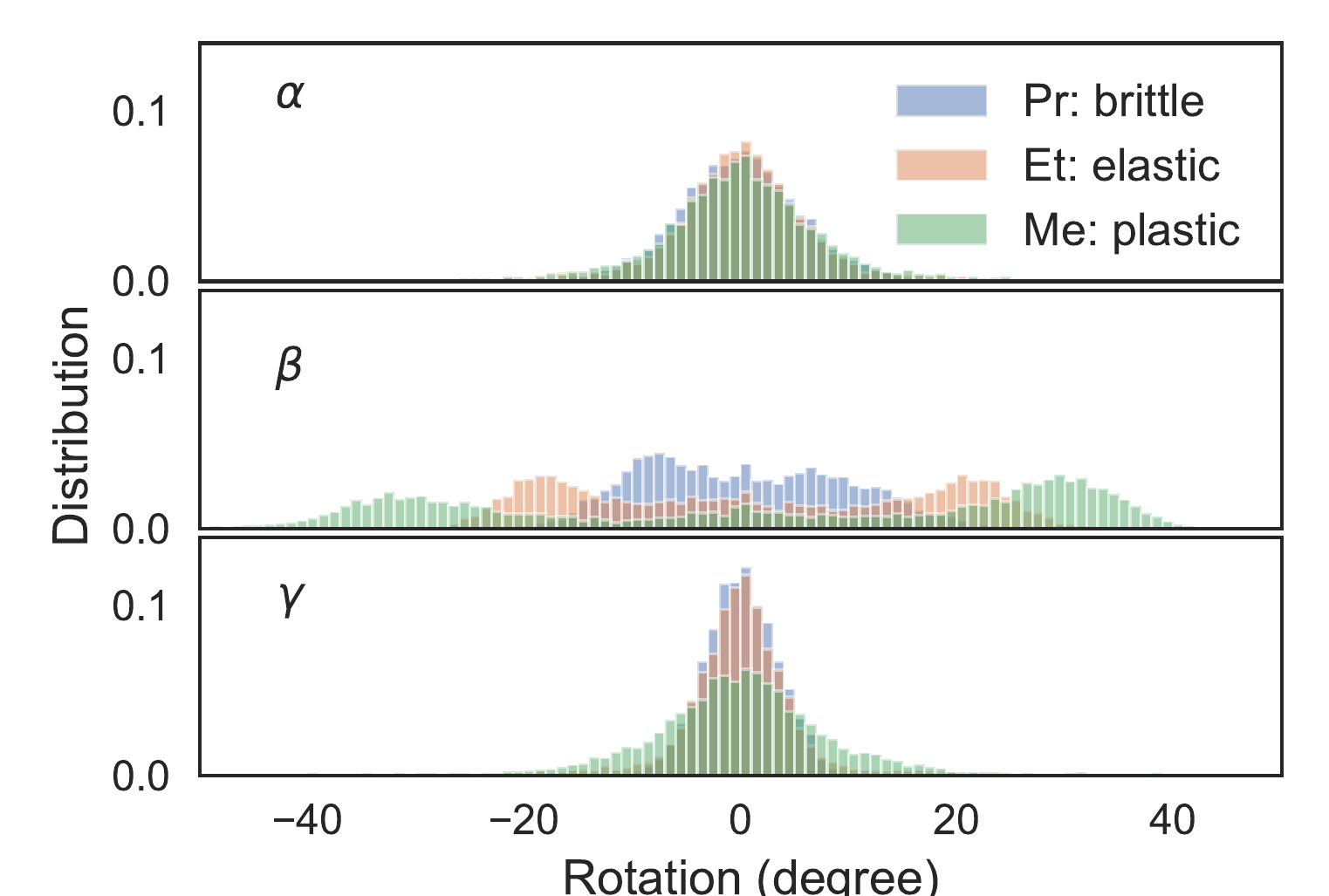}
\caption{\label{Fig5} The simulated distribution of accumulated rotational angles (with respect to the initial configurations) for all materials upon the bending loads (3.6 nm for \textbf{Pr}, 6.2 nm for \textbf{Et} and 10.1 nm for \textbf{Me}).}
\end{figure}

\begin{figure*}[htbp]
\centering
\includegraphics[width=1.0 \textwidth]{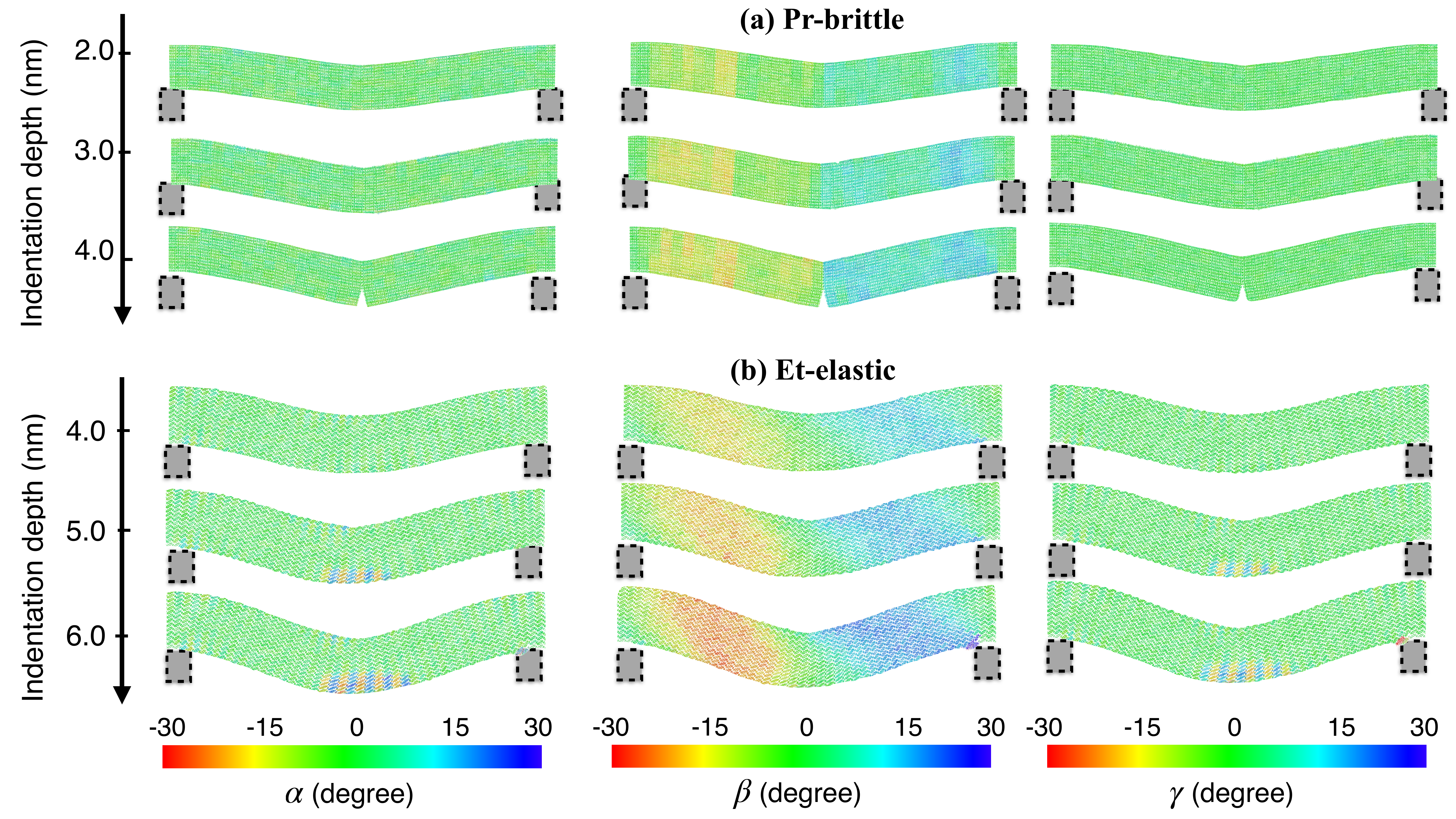}
\caption{\label{Fig6} The list of representative snapshots from the simulations of (a) \textbf{Pr}-brittle and (b) \textbf{Et}-elastic deformations.}
\end{figure*}

\subsection{Atomistic motions upon the deformation}
While analyzing their dynamic trajectories, we observed that molecules rotate strongly upon bending. Fig. \ref{Fig1} defines the alignments ($\alpha, \beta, \gamma$) for each molecule that can rotate along the $x, y, z$ axes in the Cartesian coordinates. The distributions of molecular rotations under bending are shown in Fig. \ref{Fig5}. Given that indentation direction acts on the $z$-axis and the setup of three bending points aligns along the $x$-axis, we expect that the rotation along $y$ axis ($\beta$) is the primary motion under the loading. Indeed, Fig. \ref{Fig5} reveals that the rotation in $\beta$ is more pronounced that other directions. 



\begin{figure*}[htbp]
\centering
\includegraphics[width=1.0 \textwidth]{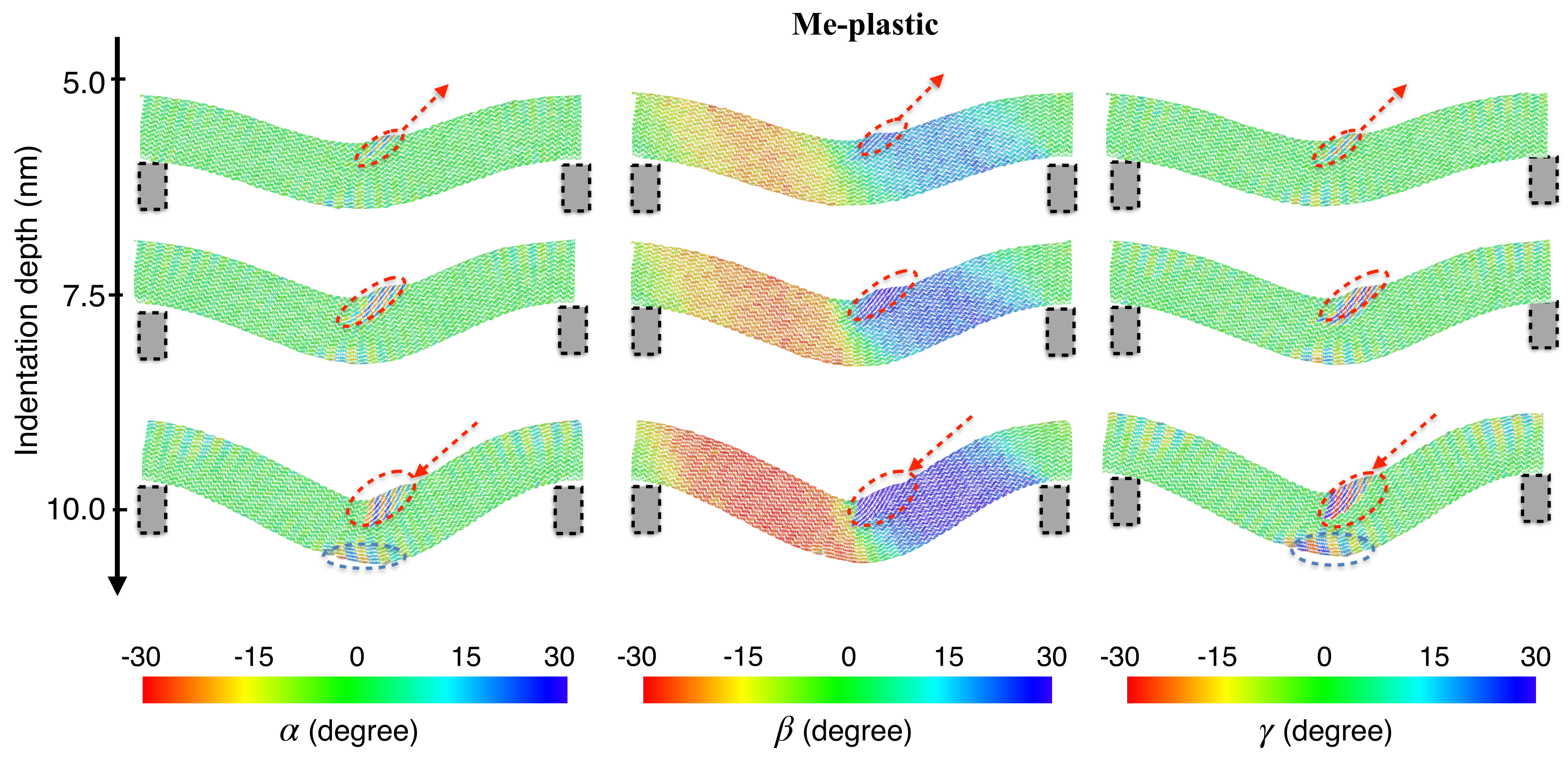}
\caption{\label{Fig7} The list of representative snapshots from the simulation of \textbf{Me}-plastic deformation.}
\end{figure*}

To understand the role of molecular rotation in the whole deformation process, we plotted a few representative structures from the MD trajectory for each system in Fig. \ref{Fig6} and Fig. \ref{Fig7} and analyzed their patterns as follows.

\textbf{Pr-brittle.} Upon deformation, we found that the sample continuously to bend from 0 to 2.5 nm (the first row of Fig. \ref{Fig6}a) and 3.5 nm (the second row of Fig. \ref{Fig6}a). The \textbf{Pr} molecules barely rotate around $x$ and $z$ axis. However, the rotation on $y$-axis is more pronounced and it symmetrically distributed around the central indenter. When the indentation depth exceeds 4.2 nm (the last row of Fig. \ref{Fig6}a), the lower surface cracks due to a large tensile stress. 

\textbf{Et-elastic.} Upto the indentation depth of 4.0 nm (the first row of Fig. \ref{Fig6}b), the \textbf{Et} molecules barely rotate around the $x$ and $z$ axis, while the rotation on $y$-axes ($\beta$) is more pronounced and it symmetrically distributed around the central indenter. However, it is clear that the molecules around the center of $y$-axis do not rotate. Upon further indentation at 5.0 nm (the second row of Fig. \ref{Fig6}b) and 6.2 nm (the last row of Fig. \ref{Fig6}b), the molecules at the center of lower surface undergo a large rotation around the $x$ and $z$ due to a large tensile stress, but do not rotate around $y$. This suggests that molecules upon tension prefer a rotation on $\alpha$ and $\gamma$, rather than the rotation around $\beta$ due to the anisotropic behavior of its potential energy landscape. Since the rotations are symmetrically distributed around the indenter, it is an elastic deformation. When the indentation is released, the process is supposed to be reversible. \textbf{Me}-Elastic sample undergoes very similar processes except that the critical indetentation depth (10.2 nm) is larger. 

\textbf{Me-plastic.} At 5.5 nm, we found that the \textbf{Me} molecules near the indenter (first row of Fig. \ref{Fig7}) have alternative changes of $\alpha$ and $\gamma$ angles, which is similar to that in Fig. \ref{Fig6}b. In addition, these molecules have non-symmetric distribution of $\beta$ angles, which signals a phase transition triggered by the large compressive stress in the upper surface due to bending. This domain of new phases, consisting of realigned molecules (denoted as the red dotted eclipse), can easily slip along its interface with the parent domain. Upon indentation, the molecules in the secondary domain do not gain enough momentum to go downward as compared to other molecules. Therefore, the relative slipping direction of the secondary domain is upward and we observed the appearance of a bump near the indenter tip (second row of Fig. \ref{Fig7}) at 6.7 nm. As the tip continues to go down, the secondary domain keeps climbing up until the bump reaches its maximum. In the mean time, the molecules at the center bottom region are nearly flattened, which can trigger another phase transition to form a new phase domain. Upon further compression, the flattened molecules at the center bottom region create much empty space along the $z$-axis. Thus, the secondary domain slips down to push the neighboring molecules down to fill the empty space (third row of Fig. \ref{Fig7}) at 9.5 nm. When the indentation is released, the process is supposed to be irreversible at low temperature since triggering the back transformation requires some energy barrier. Therefore, it is a plastic deformation. 

Clearly, such a plastic deformation process is driven by the molecular rotation, which is different from metal's plastic deformation that requires the migration of dislocations \cite{li2018sample, Zhang_2014, NOHRING2016140, ZHUO2018331, KATAKAM2020106674, HE20223687}. In several recent experimental studies, it has been proposed that molecular rotation may play a central role to generate a crystal twining \cite{takamizawa2018superplasticity} or phase transition \cite{Takamizawa2013Superelastic, karothu2016shape} which leads to plastic deformability. Our simulation on \textbf{Me}-plastic revealed a similar atomistic picture except that its new domain size is much smaller. Due to molecular rotation, some \textbf{Me} molecules near the indenter form a new phase. The newly formed secondary phase can freely slide along the interface to adjust the local stress. In the early stage, the upward movement of re-aligned molecules results in a bump shape near the indenter (instead of two bumps being symmetrically aligned near the indenter). Such an asymmetric bump has actually been found in the bending experiment \cite{devarapalli2019remarkably}, which may provide another evidence to support our modeling results. Given that most of the previous bending experiments did not report the finding of new domains, it is likely that only very small domains of rotated molecules can be formed due to energetic reasons under the plastic bending deformation. In this case, a reliable atomstic modelling is needed to capture such subtle details. Furthermore, if the temperature is sufficiently high to cross the phase transition barrier, the process may become reversible, similar to the previously reported superelastic or shape-memory phenomenon \cite{Takamizawa2013Superelastic, karothu2016shape, takamizawa2018superplasticity}.

\subsection{Rotation-dependent Energy Map}
So far, we have established the relation between molecular rotation and the observed mechanical bending flexibility of organic crystals. However, we are still unclear why some materials are more compliant than others and why we observed two different deformation behaviors on the \textbf{Me} crystal with slightly different initial configurations. To quest their physical origins, it is necessary to examine the potential energy surface (PES) with respect to the molecular rotations. 

\begin{figure}[htbp]
\centering
\includegraphics[width=0.5 \textwidth]{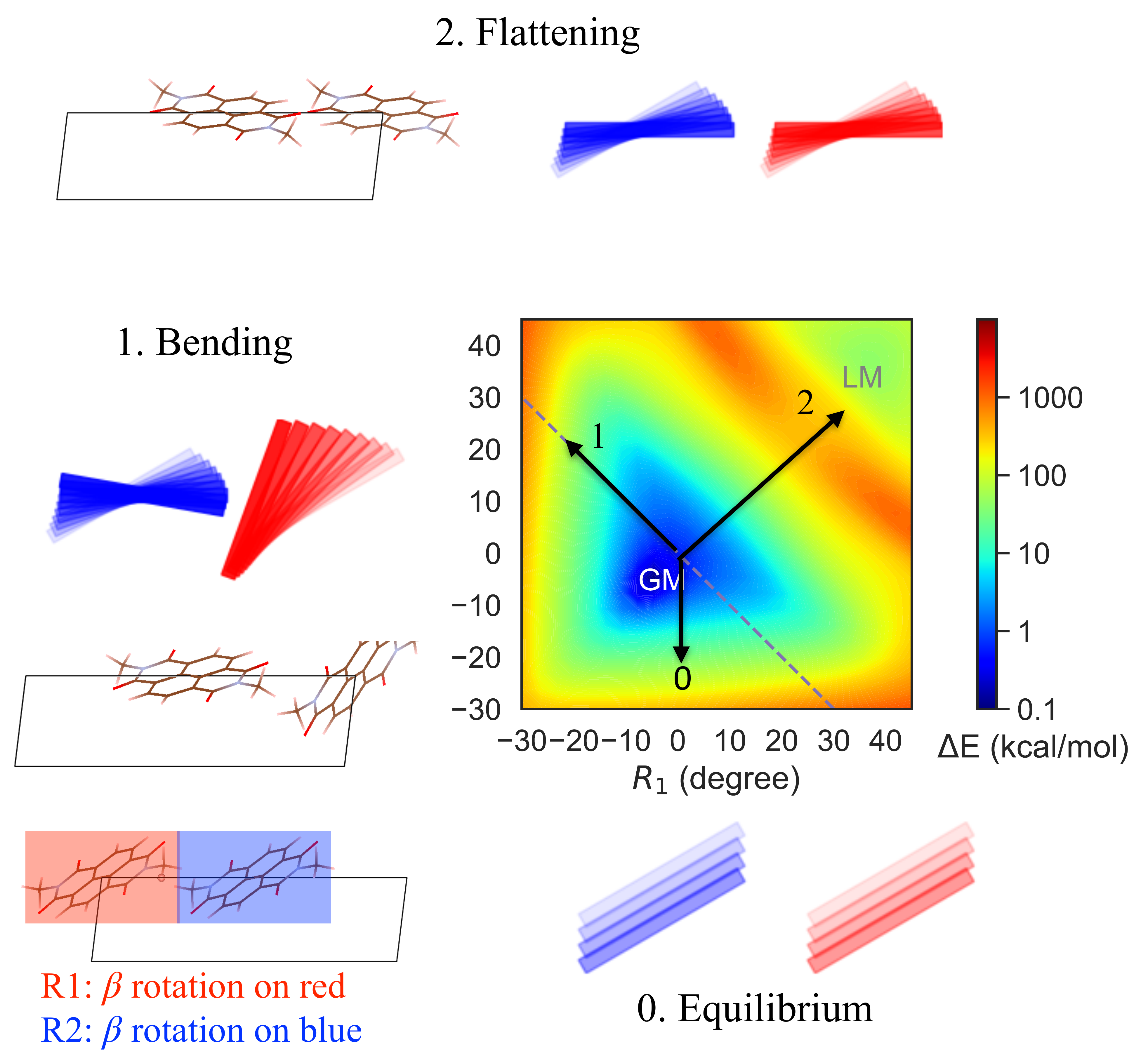}
\caption{\label{Fig8} The computed rotation-dependent energy map for Me-plastic from the GAFF model and its physical interpretation.}
\end{figure}

\begin{figure*}[!htbp]
\centering
\includegraphics[width=1.0 \textwidth]{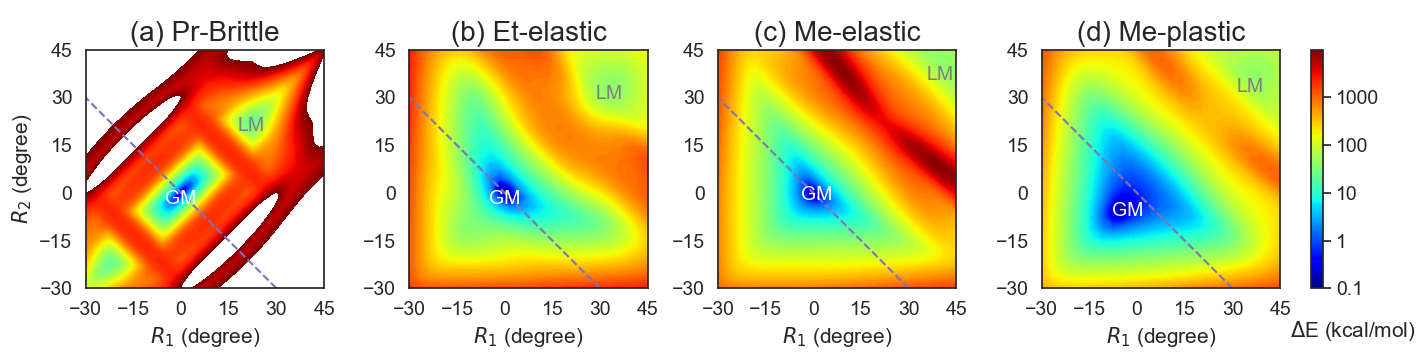}
\caption{\label{Fig9} The potential energy surface from the GAFF model as a function of molecular rotation for three crystals with different mechanical responses: (a) \textbf{Pr}-brittle, (b) \textbf{Et}-elastic, (c) \textbf{Me}-elastic, and (d) \textbf{Me}-plastic deformations. The dashed lines in each subplot denote direction of symmetric bending. The while region in (a) denotes the rotations leading to energy exceeding 10000 kcal/mol.}
\vspace{3mm}
\end{figure*}

To compute the rotation-dependent energy map, we started with the perfect crystal structures and tracked the energy changes while systematically rotating two groups of symmetrically-related molecules (colored in red and blue in Fig. \ref{Fig2}) along the $y$-axis in the unit cell. Using \textbf{Me}-plastic as an example, we computed its energy map as the function of the rotation angles ($R_1$ and $R_2$) as shown in Fig. \ref{Fig8}. In this map, it consists of two main energy basins. The basin around (0, 0) represent the global minima (GM) configuration around the equilibrium state. In the \textbf{Me} crystal, the molecules are aligned with nonzero inclination angles. The arrow from (0, 0) to (-30, 30) represents a bending of two molecules along the opposite directions (namely, clockwise and anti-clockwise directions). On the other hand, there exists another local minimum (LM) of energy at (30, 30), which represents a flattened configuration with both molecules being aligned horizontally. Such a state can be achieved through barrier crossing by adding a large tensile or compressive strains as shown in the arrow from (0, 0) to (30, 30).

Consequently, we applied this approach to compute the rotational-dependent energy maps for all systems and attempted to find a predictive model to link the possible deformation mechanism with our atomistic simulations. The results are summarized as follows, 

\begin{itemize}
    \item \textbf{Pr} has a very stiff GM (see Fig. \ref{Fig9}a). This indicates that even a slight rotation can lead to a high energy penalty. The energy basin of GM is aligned diagonally. In this energy basin, the total energy increases over 1000 kcal/mol if two molecules bend symmetrically from (0, 0) to ($\pm10, \mp10$). Such a high energy penalty would eventually lead to the formation of crack. In addition, there is a LM centered around (20, 20). But this state is nearly inaccessible from the GM due to a high energy barrier. Hence, \textbf{Pr} has a limited rotational freedom, which is consistent with its brittle nature.
    
    \item \textbf{Et} has more spreads around the GM (Fig. \ref{Fig9}b). As shown in Fig. \ref{Fig5}, two peaks are symmetrically distributed at $\pm$20 degrees when the system reaches the elastic limit. The rotation from (0, 0) to ($\pm20, \mp20$) would lead to a penalty energy of 500 kJ/mol. Therefore, the \textbf{Et} molecules can rotate more than \textbf{Pr} before the crack event starts. Similarly, \textbf{Et} has a LM around (30, 30) with a high energy barrier. 

    \item \textbf{Me}-elastic (see Fig. \ref{Fig9}c) has a shape similar to \textbf{Et} (Fig. \ref{Fig9}b), except that it has a wider bending region as denoted by the dotted line. Similarly, it has a high energy barrier that prevents the phase transition to the adjacent LM through the flattening motion. Therefore, \textbf{Me} molecules can bend more easily than \textbf{Et}, but they cannot reach the LM state due to a high barrier.

    \item \textbf{Me}-plastic has the flattest GM basin (Fig. \ref{Fig9}d). The energy barrier of symmetric bending from (0, 0) to ($\pm30, \mp30$) is only about 500 kcal/mol. Hence the \textbf{Me} molecules can bend more easily than \textbf{Et}. More interestingly, there is a low energy pathway that connects the LM at (30, 30) to the GM basin. Under the bending deformation, the molecules in a large non-periodic supercell may access other states due to the thermal fluctuation. The required barrier crossing from GM to LM can be further reduced due to the surface molecules, strain and other defects. Hence, it is possible to trigger the nucleation of a secondary domain with re-aligned molecules in the LM state. According to the nature of bending, such phase transition is more likely occur in either the upper or lower surface due to extra tensile/compressive strains. And the reoriented molecules (near the LM state) result in a stronger peak around $\beta$ = 30$^\circ$ as compared to that around -30$^\circ$ in Fig. \ref{Fig5}.   
\end{itemize}

From the above analysis, it is clear that each type of deformation has its own characteristics in their rotation-dependent energy maps despite that the model is restricted to single unit cell assumption. First, a brittle deformation should correspond to a stiff GM with high energy penalty to bend. When the GM becomes less stiffer, the system tends to have more elastic region and becomes more compliant. Finally, the key to achieve a plastic deformation is to have a low energy barrier between the GM and its adjacent LM states. Clearly, such a simplified energy map is instructive to understand the trend of bending deformation when molecular rotation is the major factor. In addition, we checked the rotation-dependent energy maps with two other energy models (OpenFF and DFTB-TS). Encouragingly, the results are overall very similar (see extended analysis and discussion in the Appendix \ref{a1}). Hence, we may be able to employ this model to predict the deformation behaviors for new organic crystals without performing expensive large scale MD simulations in the future work.

\section{Concluding Remarks}
In this work, we have performed extensive molecular dynamics simulations to directly model the mechanical bending of organic crystals. Using three recently reported naphthalene diimide derivatives as the examples, our simulation successfully produced different deformation behaviors from brittle fracture to elastic/plastic deformation upon mechanical bending. By analyzing the atomistic trajectories from our simulations, we discovered that molecular rotational freedom is the key factor determining a material's bendability, which arises from the delicate interplay between geometric packing and intermolecular interactions. Furthermore, we found the rotation-dependent potential energy surface can be used to clarify the origin of different mechanical deformation for organic materials. Although the role of molecular rotation in driving the plastic bending have been recognized in several previous experiments based on the observation of twin formation and phase transitions \cite{Takamizawa2013Superelastic, karothu2016shape, takamizawa2018superplasticity}, our work extends this mechanism to more general cases in which the rotated molecules do not necessarily form a large domain to allow the plastic deformability.

While we focused on only three naphthalene diimide derivatives in this study, the proposed three-point bending setup is entirely general to handle different organic systems as long as the crystal structures and orientations are known. In future, we will continue to test this approach on other crystallographic directions \cite{devarapalli2019remarkably} and other systems \cite{reddy2005sorting, raju2018rationalizing, zhang2021structural, panda2015spatially, reddy2005structural, saha2018molecules}. Additionally, the impacts of model size, strain rate and indenter shape on other systems need to be studied to ensure the simulation pipeline is transferable to other systems.

In parallel to this work, we recently proposed a crystal packing similarity model \cite{zhu2022quantification} that can rapidly identify the organic crystals with similar packing and intermolecular interaction. Combining it with the present atomistic modelling approach, we hope to develop a full simulation pipeline to screen new mechanically flexible organic crystals from the available database\cite{csd} for future device applications.

\begin{acknowledgements}
This research is sponsored by the NSF (DMR-2142570) and Sony Group Corporation. The computing resources are provided by ACCESS (TG-DMR180040). The authors also thank Changquan Sun, Reddy Malla, Pance Naumov, Liang Li and Sinisa Mesarovic for helpful discussions.
\end{acknowledgements}

\section*{Code availability}
The codes used to calculate the results of this study are available in \url{https://github.com/MaterSim/OST}.

\appendix
\section{Validation with Other Energy Models}\label{a1}
As discussed in the main text, the GAFF model, as well as other energy models, can describe the equilibrium lattice constants reasonably well. Since our simulations also involve samples with large deformation, it is necessary to validate the feasibility of GAFF in describing the configurations with large deformation. Hence, we performed additional validations from the following aspects.

First, we repeated the same bending simulations with the OpenFF model and the results are qualitatively similar to the simulations based on GAFF. Namely, we observed the same behaviors of brittle fracture, elastic and plastic deformations for the three systems. Given that the GAFF and OpenFF models are parameterized from completely different protocols, the agreement from two independent FF parameters indicates the observed phenomena should be general and invariant with the choice of force field models.

\begin{figure*}[htbp]
\centering
\vspace{-3mm}
\includegraphics[width=0.95 \textwidth]{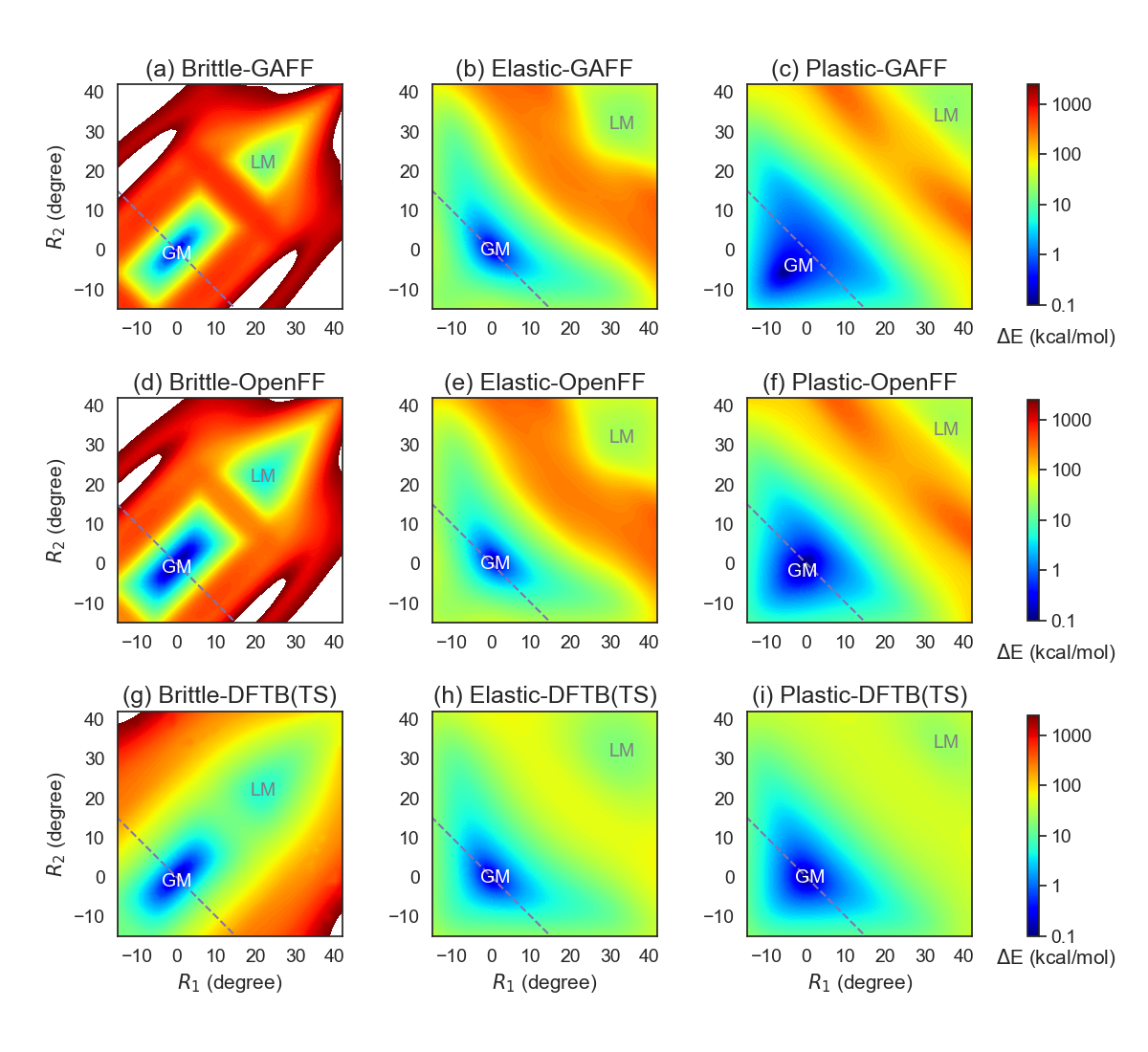}
\vspace{-5mm}
\caption{\label{Fig10} The comparison of rotation-dependent energy maps from different models.}
\end{figure*}

Second, we proposed the idea of rotation-dependent energy maps to understand the atomistic mechanism of bending in the main text. Fig. \ref{Fig10} displays the comparison of rotation-dependent energy maps from different models, including GAFF (upper panels), OpenFF (middle panels), and DFTB-TS (lower panels). It can be clearly seen that the GAFF results are remarkably similar to the OpenFF results for all three systems. Due to the convergence issue in self-consistent field method of DFTB, we omitted the configurations with rotations smaller than -15$^\circ$. The DFTB-TS approach, as a more first-principle model, also yields consistent GM and LM shapes for each system, despite that the overall energy surfaces are much smoother. If we compare the GM-LM transitions, the elastic system generally requires higher energy barrier as compared to the plastic systems for all three energy models, thus confirming our interpretation that the nucleation of LM in the plastic system is easier due to a smaller energy barrier. There is only one notable difference that DFTB-TS predict that the GM-LM transition in the brittle system requires comparable barrier than that in the plastic system. However, such transition should be prevented by the interlocking molecular packing. Hence, it does not impact our main conclusion that the transition between LM and GM in the plastic system is most favorable. 

\bibliography{ref}

\end{document}